\documentclass[journal=nalefd,manuscript=letter,layout=twocolumn]{achemso}
\usepackage[english]{babel}
\usepackage{graphicx,epsfig}
\usepackage{amsmath, amssymb, latexsym}
\usepackage{color}
\usepackage{colortbl}

\author{Louis Veyrat}
\email{l.veyrat@ifw-dresden.de}
\affiliation{IFW-Dresden, Institute for Solid State Research, PF 270116, D-01171 Dresden, Germany}
\author{Fabrice Iacovella}
\affiliation{Laboratoire National des Champs Magn\'{e}tiques Intenses (LNCMI-EMFL), UPR 3228, CNRS-UJF-UPS-INSA, 143 Avenue de Rangueil, 31400 Toulouse, France}
\alsoaffiliation{Foundation for Research and Technology, IESL, PO Box 1385, 71110 
Heraklion, Crete, Greece}
\alsoaffiliation{Department of Physics, University of Crete, 70013, Crete, Greece}
\author{Joseph Dufouleur}
\affiliation{IFW-Dresden, Institute for Solid State Research, PF 270116, D-01171 Dresden, Germany}
\author{Christian Nowka}
\affiliation{IFW-Dresden, Institute for Solid State Research, PF 270116, D-01171 Dresden, Germany}\author{Hannes Funke}
\affiliation{IFW-Dresden, Institute for Solid State Research, PF 270116, D-01171 Dresden, Germany}\author{Ming Yang}
\affiliation{Laboratoire National des Champs Magn\'{e}tiques Intenses (LNCMI-EMFL), UPR 3228, CNRS-UJF-UPS-INSA, 143 Avenue de Rangueil, 31400 Toulouse, France}
\author{Walter Escoffier}
\affiliation{Laboratoire National des Champs Magn\'{e}tiques Intenses (LNCMI-EMFL), UPR 3228, CNRS-UJF-UPS-INSA, 143 Avenue de Rangueil, 31400 Toulouse, France}
\author{Michel Goiran}
\affiliation{Laboratoire National des Champs Magn\'{e}tiques Intenses (LNCMI-EMFL), UPR 3228, CNRS-UJF-UPS-INSA, 143 Avenue de Rangueil, 31400 Toulouse, France}
\author{Bernd B\"uchner}
\affiliation{IFW-Dresden, Institute for Solid State Research, PF 270116, D-01171 Dresden, Germany}\author{Silke Hampel}
\affiliation{IFW-Dresden, Institute for Solid State Research, PF 270116, D-01171 Dresden, Germany}\author{Romain Giraud}
\affiliation{IFW-Dresden, Institute for Solid State Research, PF 270116, D-01171 Dresden, Germany}
\alsoaffiliation{Laboratoire de Photonique et Nanostructures, 91460 Marcoussis, France}

\title{Band bending inversion in Bi$_2$Se$_3$ nanostructures}
\begin{document}

\begin{tocentry}
\includegraphics[width=9cm]{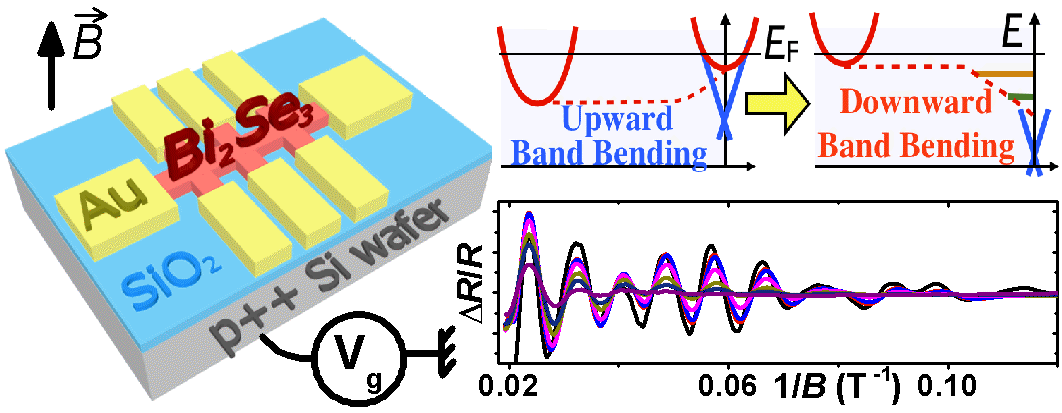}
\end{tocentry}

\begin{abstract}
Shubnikov-de-Haas oscillations were studied under high magnetic field in Bi$_2$Se$_3$ nanostructures grown by Chemical Vapor Transport, for different bulk carrier densities ranging from $3\times10^{19}\text{cm}^{-3}$ to $6\times10^{17}\text{cm}^{-3}$.
The contribution of topological surface states to electrical transport can be identified and separated from bulk carriers and massive two-dimensional electron gas.
Band bending is investigated, and a crossover from upward to downward band bending is found at low bulk density, as a result of a competition between bulk and interface doping.
These results highlight the need to control electrical doping both in the bulk and at interfaces in order to study only topological surface states. \\
\end{abstract}

A well-known limitation to the study of metallic surface states in many 3D topological insulators is their finite bulk conductivity\cite{Brahlek2014, Analytis2010}. This residual doping often originates from native defects in the crystal. In Bi$_2$Se$_3$\cite{Xia2009, Hsieh2009}, electrical doping indeed happens through
thermodynamically stable Se vacancies acting as double donors\cite{Horak1990,Xue2013}, inducing a residual bulk carrier density\cite{Analytis2010,Horak1990} which can be as large as $10^{19}\text{cm}^{-3}$. Therefore, the full conductance results from the contributions of both bulk and topological surface states (TSS). Studying the TSS then requires to identify the their own contribution to charge transport.
In Bi$_2$Se$_3$ nanostructures, this was proved possible by studying quantum transport: Aharonov-Bohm effect\cite{Peng2009, Dufouleur2013}, Conductance Fluctuations\cite{Dufouleur2015}, and Shubnikov-de-Haas oscillations (SdHO)\cite{Petrushevsky2012,Analytis2010b,Fang2012,Cao2012,Jauregui2015,Yan2014,Analytis2010,Qu2013,Ren2011,Zhang2014,Sacepe2011}.
So far, the analysis of SdHO in Bi$_2$Se$_3$ remains unclear. Some studies reported on single-band SdHO\cite{Petrushevsky2012,Analytis2010b,Fang2012,Cao2012,Jauregui2015,Yan2014}, which were attributed to either bulk carriers or TSS. In contrast, other studies reported on multi-band SdHO\cite{Analytis2010,Qu2013,Ren2011,Zhang2014,Sacepe2011} and emphasized the difficulty to identify and distinguish between their bulk, TSS, or charge-accumulation two-dimensional electron gas (2DEG) origin. In this context, an important aspect which must be considered is the band bending (BB) occuring at the surface/interface due to charge transfer\cite{Brahlek2014}. It was investigated by ARPES\cite{Bianchi2010,Benia2011,King2011,Bianchi2011,Bianchi2012,ViolBarbosa2013} and transport measurements\cite{Brahlek2014a, Analytis2010, Analytis2010b}, with seemingly contrasted reports of upward (UBB)\cite{Brahlek2014a, Analytis2010, Analytis2010b,Wolos2012} or downward (DBB)\cite{Bianchi2010, Benia2011, King2011, Bianchi2011, Bianchi2012,ViolBarbosa2013} band bending. Yet,  this effect is mostly overlooked in the analysis of SdHO, and a systematic study of both multi-band transport and band bending in Bi$_2$Se$_3$ nanostructures is still missing.
\\ In this letter, we study Shubnikov-de-Haas oscillations in nanostructures of Bi$_2$Se$_3$ under high magnetic fields for different bulk carrier densities ranging over two orders of magnitude, from $3\times10^{19}\text{cm}^{-3}$ down to $6\times10^{17}\text{cm}^{-3}$. All contributions to charge transport are separated and identified by their temperature and electrical back-gate voltage dependence. Band bending is systematically investigated, and a crossover from UBB to DBB is evidenced when bulk doping is reduced. Remarkably, a strong DBB induced by the SiO$_2$ substrate results in the formation of a potential well at the interface where a confined 2DEG coexists with TSS. This study underlines the importance of controlling the surfaces quality, in addition to solely reducing the bulk residual doping, in order to study only the TSS.
\begin{figure*}[t]
\includegraphics[width=17.78cm]{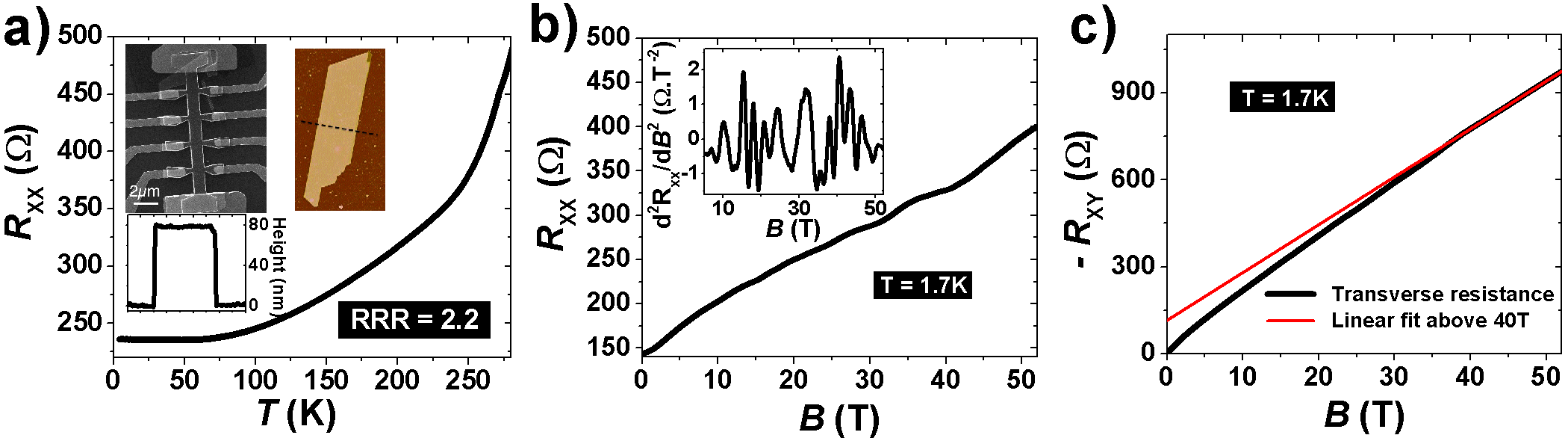}
\caption{a) Temperature dependence of longitudinal resistance $R_\text{xx}$. Insert: SEM and AFM picture of sample E, and AFM section along the indicated line. b) Perpendicular field dependence of $R_\text{xx}$ of sample E; inset: second derivative of $R_\text{xx}$ displaying SdHO. c) Perpendicular field dependence of the Hall resistance $R_\text{xy}$ of sample E ($n_\text{Hall} = 3.2\pm0.1\times10^{13}\text{cm}^{-2}$ is the total density determined from high field asymptote above 40T).}
\end{figure*}
\\
The samples studied in this paper are monocrystalline Bi$_2$Se$_3$ nanostructures grown by Chemical Vapor Transport (for growth details see \cite{Nowka2015}). Cr/Au contacts were prepared using standard e-beam lithography and metal lift-off, following an in-situ smooth Ar-ion etching to achieve good ohmic contacts. All samples show a metallic behavior, as expected for degenerate Bi$_2$Se$_3$\cite{Brahlek2014} (see fig.1.a). Five samples were studied, with different bulk doping: high density (A: $1.4\times10^{19}\text{cm}^{-3}$; B: $2.6\times10^{19}\text{cm}^{-3}$; C: $1.4\times10^{19}\text{cm}^{-3}$), intermediate density (D: $4.3\times10^{18}\text{cm}^{-3}$) or low density (E: $6.6\times10^{17}\text{cm}^{-3}$) (see table 1). 
\begin{figure}[H]
\includegraphics[width=8.46cm]{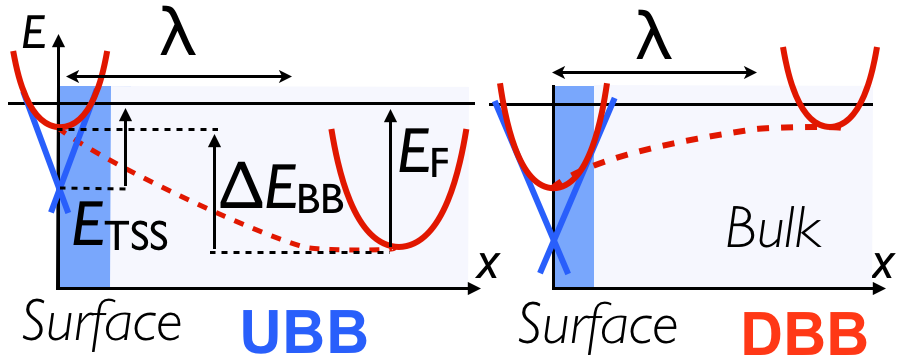}
\caption{Schematics of BB: left, UBB with $\lambda$ the depletion length and $\Delta E_\text{BB}>0$ the band bending energy; right, DBB.}
\end{figure}
For each structure, magneto-transport was studied down to 4K and up to 55T using a pulsed magnetic field, except for structure A which was studied under a static field up to 15T. From Hall bar measurements, the longitudinal resistance $R_\text{xx}$ shows a weak amplitude SdHO superimposed to a large magnetoresistance background (see fig.1.b) and a non-linear transverse resistance $R_\text{xy}$, which is typical of multiband transport (see fig.1.c). More details on the different contributions to electronic transport properties can be inferred from SdHO, which quantitative analysis is better performed on the second derivative $\text{d}^2R_\text{xx}/\text{d}B^2$.
A Fast Fourier Transform (FFT) analysis was performed on SdHO to extract the frequencies corresponding to the different electronic populations. In order to identify them, the temperature and back-gate voltage dependences were studied. Indeed, bulk carriers and TSS can be told apart through their effective mass $m^*$ \cite{Petrushevsky2012, Analytis2010}, which is extracted from the temperature dependence of SdHO using the Lifschitz-Kosewitz (LK) formula\cite{Schoenberg1984}. Moreover, the interface TSS (located at the interface between Bi$_2$Se$_3$ and SiO$_2$) is gate sensitive, contrary to bulk states and surface TSS (TSS at the interface with vacuum)\cite{Sacepe2011}. From the SdH frequency associated with each population, the carrier density, the Fermi energy and the band bending $\Delta E_\text{BB} = E_\text{F} - (E_\text{surf/int}-180\text{meV})$ are extracted (see fig.2), assuming that the Dirac point is 180meV below the bottom of the conduction band (CB)\cite{Xia2009, Hsieh2009}.
\\
A typical strongly-doped ribbon (sample A) is shown in figure 3.a. Above 6T, the resistance exhibits SdHO, that are visible up to 15T (see fig.3.b). Those oscillations show a clear multifrequency pattern: from $\sim$11T on, the oscillations display a third frequency significantly higher than below 11T. The FFT reveals three main peaks (similarly to earlier results\cite{Qu2013}), at $f_\text{B1} = 165\pm5\text{T}$, $f_\text{B2} = 180\pm5\text{T}$ and $f_\text{B3} = 245\pm5\text{T}$, which correspond to three distinct electronic populations having different carrier densities. In order to identify each contribution, SdHO were measured at a back-gate voltage of $V_\text{g}$ = 0 and -110V. The Fourier transforms are compared in figure 3.c. As pointed out by the dotted lines, both $f_\text{B2}$ and $f_\text{B3}$ do not shift at -110V, while $f_\text{B1}$ is clearly shifted to a lower frequency. This indicates that $f_\text{B1}$ corresponds to the interface TSS.
Hence, we calculate: $n_\text{int} = f_\text{B1}\times\left(e/h\right) = 4.0\pm0,1\times10^{12}\text{cm}^{-2}$, with $e$ the electronic charge and $h$ the Planck constant. Knowing the Fermi velocity $v_\text{F} = 5.4\times10^{5}\text{m.s}^{-1}$ determined by ARPES\cite{Kuroda2010, Xia2009, Hsieh2009}, the interface Fermi energy is $E_\text{int} = 251\pm3\text{meV}$ and $m^*_\text{int} = E_\text{int}/v_\text{F}^2 = 0.15 m_\text{e}$ is the associated cyclotron mass. The two other frequencies can be separated by their temperature dependences. Since the onset of the highest frequency $f_\text{B3}$ is 11T, the temperature dependence of the SdHO amplitude below 11T arises only from the two populations associated to $f_\text{B1}$ and $f_\text{B2}$. Although no quantitative value of $m^*$ for one single population can be extracted, LK fits for SdHO peaks below 11T yield an averaged effective mass $m^* = 0.14\pm 0.01 m_\text{e}$ (see fig.3.d and Supplementary materials). This is consistent with $f_\text{B2}$ originating from bulk carriers. Indeed, $m^*_\text{B} = 0.12-0.15m_\text{e}$ from bulk and $m^*_\text{int}$ from interface are very similar, and temperature dependences are similar for both populations. Above 11T, $m^*$ extracted from LK-fits rises to $0.19\pm 0.02 m_\text{e}$. This is in agreement with the last frequency $f_\text{B3}$ having a higher effective mass than the previous two. Therefore we attribute $f_\text{B3}$ to the surface TSS, with an energy $E_\text{surf} = 305\pm3\text{meV}$ and a corresponding effective mass $m^*_\text{surf} = 0.185 m_\text{e}$, in agreement with the LK fits.
This is also consistent with the difference found in the onsets of the SdHO, which depend on the carrier mobility of each band. Indeed, the Drude mobility $\mu = e \tau_\text{e}/m^*$ depends on the disorder strength (characterized by $\tau_\text{e} = l_\text{e}/v_\text{F}$, where $l_\text{e}$ is the elastic mean free path) and on the effective mass\cite{Brahlek2014}. Here, the carrier density is 50$\%$ higher at the surface than at the interface. As doping mostly originates from double donor Se vacancies\cite{Devidas2014,Xue2013}, a higher doping means a stronger disorder, and a reduced $l_\text{e}$. Simultaneously, for TSS, increasing energy implies increasing $m^*$. Both effects tend to reduce the surface TSS' mobility compared to the bulk and interface one.
\\ 
\begin{figure*}[t!]
\includegraphics[width=17.78cm]{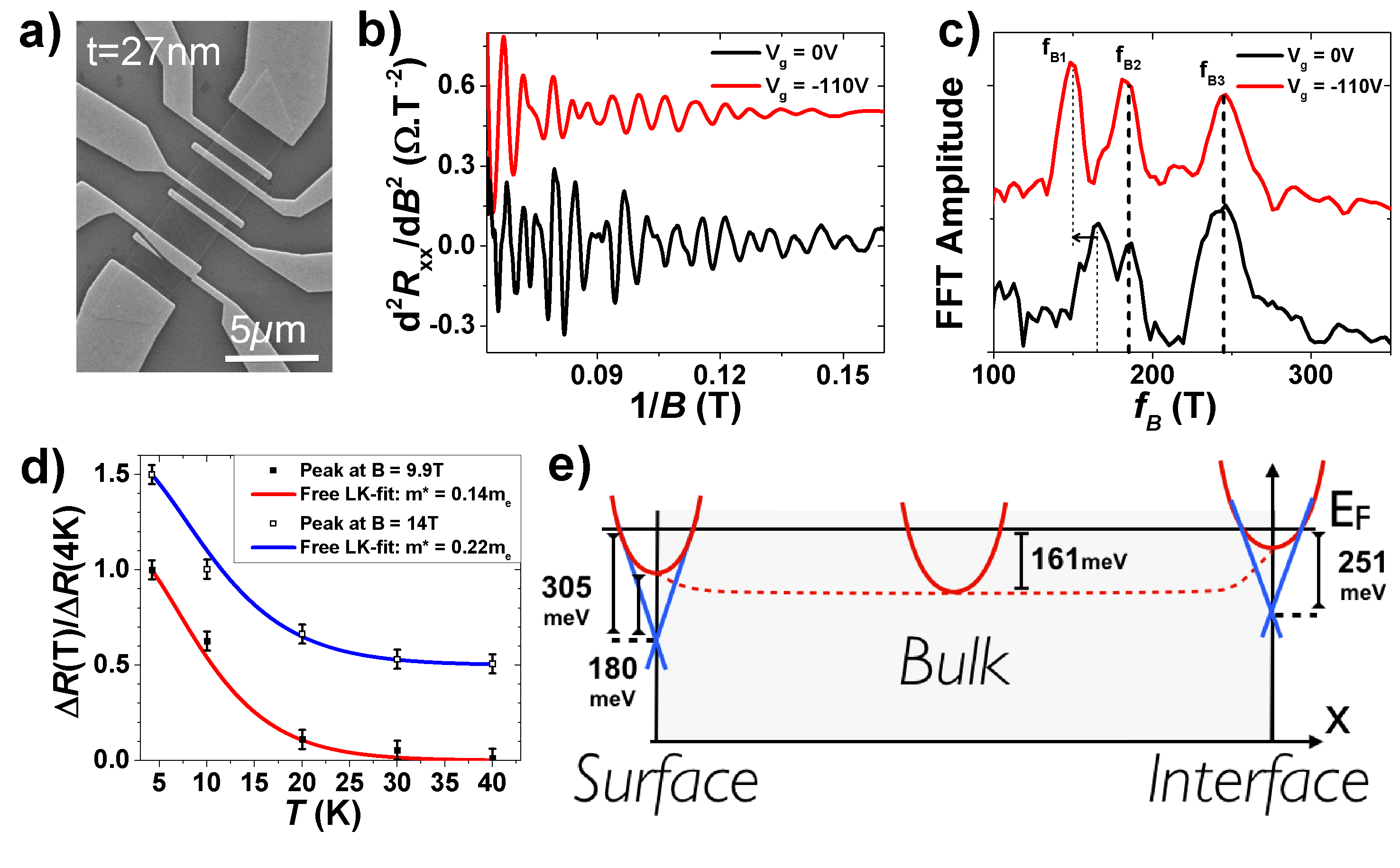}
\caption{Magneto-transport study of strongly doped sample A (thickness t=27nm): a) SEM picture of the contacted sample A; b) 2$^{nd}$ derivative of $R_\text{xx}$ against inverse magnetic field, at two gate voltages; c) FFT of SdHO at different gate voltages (curve at $V_\text{g}$ = -110V shifted for clarity); d) Temperature dependance of the amplitude of some SdH peaks (points at 13.2T shifted for clarity); e) Schematic of the band bending in sample A.}
\end{figure*}
Assuming a quadratic energy dispersion for the bulk, the BB can be calculated as shown in figure 3.e. An UBB is observed in this structure, as expected for a strongly doped structure ($n_\text{B} = 1.4\times10^{19}\text{cm}^{-3}$). 
Indeed, if surface doping only originates from bulk charge transfer and not from additional surface doping, the charges from the bulk are transferred to the surface over a depletion length $\lambda$, resulting in UBB. Here, $\lambda > 3-4\text{nm}$ assuming a complete depletion of the bulk over this range. $\lambda$ can be increased by decreasing the bulk carrier density; ultimately, for low enough doping and for nanostructures thinner than 2$\lambda$, the bulk could be fully depleted\cite{Brahlek2014}.
\\ \\
Three other samples (B,C,D), with comparable or lower doping, were studied under magnetic field up to 55T. As for sample A, the FFTs of the multi-frequency SdHO exhibits three peaks, associated with bulk and two TSS (see Supplementary materials). Because of the absence of a back-gate voltage on these structures, surface and interface TSS cannot be told apart. Still, all those samples show UBB, as can be seen in table 1. For all these structures, $\Delta E_\text{BB}$ is of the same order of magnitude ($\sim$50 to $\sim$150meV); but the effect, small in strongly doped structures (for A,B,C: $E_\text{F}$ lies in the CB at the surfaces), becomes much stronger for the intermediate doping (sample D), where $E_\text{F}$ enters the bulk gap for one surface ($E_\text{surf2} <$ 180meV). This confirms the possibility, for thin enough structures with doping in the $10^{18}\text{cm}^{-3}$ range, to achieve bulk depletion through UBB. However, for this doping, the depletion length $\lambda$ would be around 3nm for a complete bulk depletion and 6nm for a triangular depletion potential, so that bulk depletion can only be achieved for a thickness below 12nm, which constitutes a challenging task nowadays. To achieve bulk depletion over a larger thickness would require to further decrease the bulk carrier density. However, this assertion supposes that BB remains upward, which is not obvious, as discussed below.
\\
\begin{table*}[botcap]
\begin{tabular}{ | l | p{1.3cm} | p{1.2cm} | p{1.9cm} || p{1.3cm} | p{1.1cm} || p{1.3cm} | p{1.2cm} | p{2.1cm} |}
    \hline
    Sample & $n_\text{surf}$ (cm$^{-2}$) & $E_\text{surf}$ (meV) & $\Delta E_\text{BB-surf}$ (meV) & $n_\text{B}$ (cm$^{-3}$) & $E_\text{F}$ (meV) & $n_\text{int}$ (cm$^{-2}$) & $E_\text{int}$ (meV) & $\Delta E_\text{BB-int}$ (meV)\\ \hline
    A & 5.9x10$^{12}$ & 305 & +36 \textcolor{blue}{\textbf{UBB}} & 1.4x10$^{19}$ & 161 & 4x10$^{12}$ & 251 & +90 \textcolor{blue}{\textbf{UBB}}\\ \hline
    E & 2.9x10$^{12}$ & 215 & -14 \textcolor{red}{\textbf{DBB}} & 6.6x10$^{17}$ & 21 & 1.3x10$^{13}$ & 459 & -258 \textcolor{red}{\textbf{DBB}}\\
    \hline
    \hline
    Sample & $n_\text{surf1}$ & $E_\text{surf1}$ & $\Delta E_\text{BB-surf1}$ & $n_\text{B}$ & $E_\text{F}$ & $n_\text{surf2}$ & $E_\text{surf2}$ & $\Delta E_\text{BB-surf2}$ \\ \hline
    B & 1.0x10$^{13}$ & 396 & +35 \textcolor{blue}{\textbf{UBB}} & 2.6x10$^{19}$ & 251 & 3.9x10$^{12}$ & 247 & +184 \textcolor{blue}{\textbf{UBB}}\\ \hline
    C & 6.3x10$^{12}$ & 317 & +38 \textcolor{blue}{\textbf{UBB}} & 1.5x10$^{19}$ & 175 & 2.7x10$^{12}$ & 206 & +149 \textcolor{blue}{\textbf{UBB}}\\ \hline
    D &  2.7x10$^{12}$ & 207 & +48 \textcolor{blue}{\textbf{UBB}}& 4.3x10$^{18}$ & 75 & 1.2x10$^{12}$ & 136 & +119 \textcolor{blue}{\textbf{UBB}}\\
    \hline
    \end{tabular}
    Table 1: Table of the carrier densities, band bending and Fermi energies for bulk and TSS, extracted from SdHO analysis. Top: devices A and E, for which the interface is identified by its back-gate voltage dependence. Bottom: devices B,C,D, measured without an electrical back gate.
    \end{table*}
\\ Magneto-transport results on the weakly-doped nanostructure E are presented in figure 4. Complex SdHO are observed in $R_\text{xx}$ (see figure 4.b), with the first oscillations observed at $\sim$6T. The multifrequency behavior arises above 15T, which imposes to work at high magnetic fields to study all populations. The FFT reveals a much more complex pattern than for the previous samples (see figure 4.c). Five main peaks are identified, at $f_\text{B1} = 24\pm3\text{T}$, $f_\text{B2} = 90\pm4\text{T}$, $f_\text{B3} = 121\pm5\text{T}$, $f_\text{B4} = 305\pm20\text{T}$ and $f_\text{B5} = 550\pm30\text{T}$. Importantly, the lowest frequency f$_{B1}$ corresponds to the population with the highest mobility, which gives a single-band behavior below 15T. The LK fit of the temperature dependence of the SdHO amplitude below 15T gives a $m^* = 0.129\pm0.01 m_\text{e}$ (see figure 4.d), which is consistent with bulk states\cite{Koehler73, Butch2010}. It is also inconsistent with TSS, as for such $f_\text{B1}$ the chemical potential would be 64meV and the corresponding cyclotron mass $m^*_\text{TSS} = 0.039 m_\text{e}$. The lowest frequency is therefore unambiguously identified as bulk states, with $E_\text{F} = 21\text{meV}$ and $n_\text{B} = 6.6\times10^{17}\text{cm}^{-3}$. The angle dependence of the SdHO, presented in Supplementary materials, confirms that all other frequencies than $f_\text{B1}$ tend to disappear when tilting the magnetic field from perpendicular to in-plane.\\
In order to identify the remaining populations, the back-gate voltage dependence of SdHO was studied up to -32V. The FFT allows us to separate the gate-sensitive ($f_\text{B4,5}$) and gate-insensitive ($f_\text{B1,2,3}$) frequencies. Out of the four 2D bands  $f_\text{B2,3,4,5}$, the only consistent model is to associate $f_\text{B3}$ and $f_\text{B5}$ to surface and interface TSS, respectively (see Supplementary materials). The topological carrier densities are then $n_\text{surf} = 2.9\pm0.1\times10^{12}\text{cm}^{-2}$ and $n_\text{int} = 1.33\pm0.07\times10^{13}\text{cm}^{-2}$, corresponding to energies $E_\text{surf} = 215\pm4\text{meV}$ and $E_\text{int} = 459\pm12\text{meV}$. As a consequence, the strong DBB near the interface induces a quantum well, into which a confined 2DEG from bulk origin may form. Modeling this well with a triangular potential, one can calculate for a thickness $\lambda = 9.9\pm0.9\text{nm}$ that two confined sub-bands will be partially filled, with energies $40\pm2$ and $143\pm8\text{meV}$. They perfectly match the remaining $f_\text{B2}$ and $f_\text{B4}$ ($40\pm2$ and $136\pm8\text{meV}$) (see fig.4.e). The thickness of the quantum well is in agreement with previously reported DBB\cite{Benia2011,King2011,Bianchi2011,Bianchi2010,ViolBarbosa2013}.
The total carrier density, computed as $n_\text{tot}=n_\text{B} \times \left(t-\lambda\right) + \sum\limits_{i=2}^{i=5} n^i_\text{2D} = 3.0\pm0.2\times10^{13} \text{cm}^{-2}$ taking all five contributions into account, is confirmed by Hall measurement ($n_\text{Hall} = 3.2\pm0.1\times10^{13}\text{cm}^{-2}$, see figure1.b).
\begin{figure*}[t]
\centering
\includegraphics[width=17.78cm]{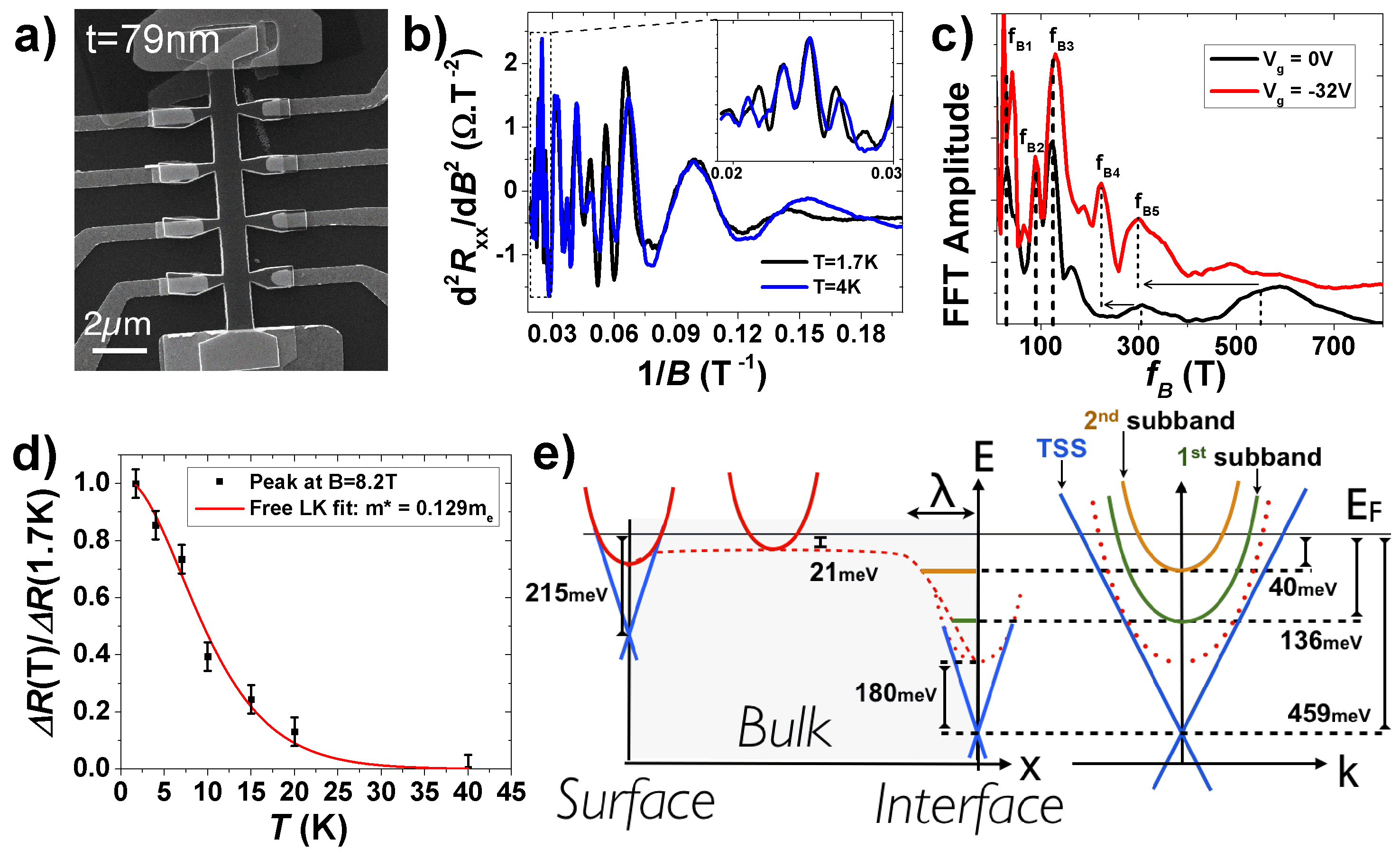}
\caption{Magneto-transport study of weakly doped sample E (t=79nm): a) SEM picture of the Hall-bar designed flake E; b) 2$^{nd}$ derivative of $R_\text{xx}$ against inverse magnetic field, at two temperatures to show reproducibility (curve at -110V in Supplementary materials); c) FFT of the SdHO at different gate voltages; d) Temperature dependance of the amplitude of a SdH peak; e) schematic of the band bending in sample E.}
\end{figure*}
\\
Although the bulk is very weakly doped, BB is here downward. This reflects the fact that the interface doping is much higher than the bulk one. Such a strong effect on the BB is only visible because of the small bulk doping. For a similar surface doping in sample B (surface 1) but with a much larger bulk carrier density, BB remains upward due to the large Fermi energy in the bulk.
This points at the necessity to control both the residual bulk doping and the additional surface/interface doping in order to produce bulk-compensated structures. In particular, the very strong DBB near the interface suggests that a SiO$_2$ substrate favors a strong interface doping, which could result either from a locally-increased disorder in Bi$_2$Se$_3$ or from chemical bonds. The use of an alternative oxygen-free substrate (GaAs, SiC, GaN...) could maybe overcome this effect.
\\
The difference in the observed mobilities is also consistent with the type of band bending and the assumption that disorder is stronger at interfaces with respect to the bulk. In sample E, the situation $\mu_\text{int} < \mu_\text{surf} < \mu_\text{b}$ is well understood by $l_\text{e-b} > l_\text{e-TSS}$ and $m^*_\text{int} > m^*_\text{surf} \sim m^*_\text{b}$, as a result of DBB. Indeed, bulk SdHO start first ($\sim6T$), whereas surface and interface SdHO begin at higher fields ($\sim14T$ and $\sim27T$, respectively, in the ratio of their cyclotron masses). On the contrary, for samples A,B,C,D, bulk states and TSS have a more similar mobility, which is in agreement with a reduced cyclotron mass in the case of UBB.
\\
\\
In conclusion, we investigated charge transport in Bi$_2$Se$_3$ nanostructures grown by CVT through a careful analysis of SdHO under high magnetic fields. This gives the possibility to identify all electronic populations, as bulk states and TSS, or even as a massive 2DEG for low bulk doping. We confirm that UBB is the usual case for strongly doped structures, due to bulk-dominated charge transfer at interfaces. Remarkably, DBB is found for a rather small bulk doping, which reveals the importance of interfaces doping and the influence of the substrate. Although UBB could indeed lead to bulk-depleted structures, our results emphasize the importance to control the interfaces quality so as to avoid downward band bending, which is necessary for further investigation and engineering of TSS transport. A possible direction maybe could be the growth on oxygen-free substrates.
\\
\\
Aknowledgements: We thank Dr. Bastien Dassonneville for helpful discussions. The support of the European Magnetic Field Laboratory (EMFL), under proposal TSC17-114, is ackowledged. J.D. acknowledges the support of the German Research Foundation DFG through the SPP 1666 Topological Insulators program.
\bibliography{paper}
\end{document}